\documentstyle[epsf,11pt]{article}
\newcommand{\bea}{\begin{eqnarray}}
\newcommand{\eea}{\end{eqnarray}}
\begin{document}
\vspace{0.2cm}

\newcommand{\lsim}
{{\;\raise0.3ex\hbox{$<$\kern-0.75em\raise-1.1ex\hbox{$\sim$}}\;}}
\newcommand{\gsim}
{{\;\raise0.3ex\hbox{$>$\kern-0.75em\raise-1.1ex\hbox{$\sim$}}\;}}

\begin{flushright}
{\large HIP-2001-29/TH \\[2mm] 
TIFR/TH/01-23} 
\end{flushright}

\begin{center}
{\LARGE\bf Single sneutrino production in $\gamma\gamma$ collisions}
\\[15mm]
{\bf M. Chaichian$^{a,b}$, K. Huitu$^a$, S. Roy$^{a,c,}$\footnote{On leave 
of absence from the Department of Theoretical Physics, TIFR, Mumbai, 
India} and Z.-H. Yu$^{a,b}$}\\[4mm]
$^a$Helsinki Institute of Physics\\[3mm]
and \\[3mm]
$^b$Department of Physics, \\
P.O.Box 64, FIN-00014 University of  Helsinki, Finland \\[4mm]
$^c$Department of Theoretical Physics, Tata Institute of Fundamental 
Research \\ Homi Bhabha Road, Mumbai - 400 005, INDIA \\[7mm]
\date{}
\end{center}

\begin{abstract}
We study the single production of sneutrinos with two leptons (or jets) 
via $\gamma\gamma$ collision in an R-parity ($R_{p}$) violating 
supersymmetric model. The subsequent decays of the sneutrino are also 
considered. The single production of sneutrinos may provide a significant 
test of supersymmetry and $R_p$-violation with flavour conserving and 
flavour changing final states. If such processes coming from $R_p$ violation 
are not detected, the parameter space of the model will be strongly 
constrained at the future Linear Collider.
\end{abstract}

\vskip 10mm

{~~~~PACS number(s): 13.65.+i, 13.88.+e, 14.65.-q, 14.80.Dq, 14.80.Gt}

\newpage

\eject
\rm
\baselineskip=0.36in

\begin{flushleft} {\bf I. Introduction} \end{flushleft}

\noindent
One of the experimentally crucial issues in supersymmetric models
is whether the R-parity  \cite{RP_Farrar} is conserved or not.
The R-parity ($R_{p}=(-1)^{3B+L+2S}$, where $B$, $L$ and $S$ 
denote the baryon number, lepton number and spin) is introduced to forbid
the fast proton decay, and it implies that the lightest supersymmetric
particle (LSP) is stable and superparticles can only be pair
produced. 
However, the conservation of $R_p$ is not a necessity \cite{RP_Hall} and it 
can be violated in many ways. Numerous new phenomena are possible, if $R_p$ 
is violated \cite{RP_Aulakh}, e.g.  $R_{p}$-violation ($\rlap/\! R_{p}$) may 
explain neutrino oscillations from atmospheric neutrinos observed in 
Super-Kamiokande \cite{neutrino}.

Searching for signals of supersymmetry (SUSY) \cite{SUSY} is one of the 
main aims of a Linear Collider (LC). LC is considered to be good at detecting 
physics beyond the Standard Model (SM) with its cleaner background in 
comparison with hadron colliders. However, its C.M. energy ($ 500 - 1000$ GeV) 
is lower than that in the future hadron colliders. Thus, if the sparticles 
are heavy, producing a single SUSY particle is kinematically favored.
In addition to the $e^{+} e^{-}$ collider mode, the LC can, with the
advent of new collider techniques, produce
highly coherent laser beams being back-scattered with high luminosity
and efficiency at the $e^{+}e^{-}$ colliders \cite{pho_tech}.
In this paper we will concentrate on $\gamma\gamma$ collisions.

The $R_p$-violation constrained by the low energy
experiments has been widely discussed \cite{RP_Phe}.
At the lepton colliders, $R_p$ violation may be detected directly
in sparticle production \cite{RP_spp} or decay \cite{RP_decay}, or
indirectly \cite{RP_indirect}.
The single production of sneutrinos from $e^+e^-$
collision has  been considered in \cite{RP_spp},
where the $L$-violating parameters $\lambda$ involving the
light flavours dominate the process. 
The resonant production of sneutrinos and single chargino production
via $\gamma\gamma$ collision has also been considered in \cite{RP_spp2},
where it was found that it is possible to improve the bounds on the
parameters of the $\rlap/\! R_{p}$ model, if SUSY with $\rlap/\!
R_{p}$ is not detected. The resonant sneutrino production at Large Hadron 
Collider (LHC) has been considered in Ref. \cite{RP_BV}. 

In this work we will consider the 
single production of scalar neutrinos accompanied by two leptons (or two 
jets) in $\gamma\gamma$ collisions,
$$
\begin{array}{l}
\gamma\gamma \rightarrow \tilde\nu + l \bar{l}^{'},\;\; 
\tilde\nu + q \bar{q}^{'}.
\end{array}
\eqno {(1.1)}
$$
Each diagram in the process contains one $R_p$-violating coupling,
see Fig. 1.
We will assume that one of the couplings dominates and thus is the
only one which needs to be considered.
Furthermore, the couplings including the third generation are less
strictly bounded, and we will consider only them in this
work. 
Thus for us the relevant couplings and the corresponding experimental
limits are (from Allanach et al in \cite{RP_Phe})
$$
\begin{array}{l}
\lambda_{131}\lsim 0.062\times \frac{m_{\tilde e_R}}{100 \;{\rm GeV}},
\;\lambda_{231}\lsim 0.070\times \frac{m_{\tilde e_R}}{100 \;{\rm GeV}},
\end{array}
\eqno {(1.2)}
$$
$$
\begin{array}{l}
\lambda^{'}_{322}\lsim 0.52\times \frac{m_{\tilde s_R}}{100 \;{\rm GeV}},
\;\lambda^{'}_{323}\lsim 0.52\times \frac{m_{\tilde b_R}}{100 \;{\rm GeV}}.
\end{array}
\eqno {(1.3)}
$$
The bounds in (1.2) are found \cite{bgh} from the measurements of
$R_\tau=\Gamma (\tau\rightarrow e\nu\bar\nu)/
\Gamma (\tau\rightarrow \mu\nu\bar\nu)$ and 
$R_{\tau\mu}=\Gamma (\tau\rightarrow \mu\nu\bar\nu)/
\Gamma (\mu\rightarrow e\nu\bar\nu)$, while
the bounds in (1.3) come from \cite{ls} 
$R_{D_s}=\Gamma (D_s\rightarrow \tau\nu_\tau)/
\Gamma (D_s\rightarrow \mu\nu_\mu)$.

The production cross section of the process in (1.1) is of similar 
magnitude than the resonant 
production cross section.  The process may induce 
flavour-changing final states. Unlike in the case of resonant production, 
we may distinguish between the $R_p$-violating sources from the 
final states, which is one of the major advantages of this channel 
compared to the resonant production.
Since Higgs single production with two leptons (or two jets) is suppressed 
by Yukawa coupling, the signal process may well have bigger cross section 
than the SM background. The background induced by Z boson can be 
distinguished with different quantum numbers and is suppressed when we
consider a proper invariant mass cut.

In section 2, the supersymmetric
$\rlap/\! R_{p}$ interactions and 
calculations of the cross sections for the processes (1.1)
are presented and detection strategy is considered.
In section 3 we present numerical calculations of the processes and possible
signals are given.
Our conclusions are given in section 4 and the expressions for the photon 
luminosity are given in the appendix.

\begin{flushleft} {\bf 2. Production and decay of $\tilde\nu$
with explicit R-parity violation}\end{flushleft}

\noindent
All renormalizable supersymmetric $\rlap/\! R_{p}$ interactions can
be introduced in the superpotential as \cite{RP_V_form}:
$$
\begin{array} {lll}
    W_{\rlap/\! R_{p}} & =\frac{1}{2}
\lambda_{[ij]k} L_{i}.L_{j}\bar{E}_{k}+\lambda^{'}_{ijk}
L_{i}.Q_{j}{\bar D_{k}}+\frac{1}{2}\lambda^{''}_{i[jk]}
\bar{U}_{i}\bar{D}_{j}\bar{D}_{k}+\epsilon _{i} L_{i} H_{u}.
\end{array}
\eqno {(2.1)}
$$
where $L_i$, $Q_i$ and $H_u$ are SU(2) doublets containing lepton, quark
and Higgs superfields respectively, $\bar{E}_j$ ($\bar{D}_j$, $\bar{U}_j$)
are the singlets of lepton (down-quark and up-quark),
and $i,j,k$ are generation indices and square brackets on them denote
antisymmetry in the bracketted indices.
We will consider only the L-violating trilinear terms in
our calculations. 
The bilinear terms \cite{bilinear} $\epsilon _{i} L_{i} H_{u}$ would also 
contribute to the process.
The additional contribution to the production of a sneutrino with
fermions comes only through mixing of Higgses and sneutrinos.
In addition to the mixing, this is suppressed by the small Yukawa
coupling, except for the top quark. When one considers the effects of 
bilinear terms in the decay of the sneutrino, several other 
possibilities exist. The processes involving the bilinear terms require
a seperate analysis altogether and we will not connect it in the context
of our present paper.

In the following calculations we assume that the parameters $\lambda$
and $\lambda^{'}$ are real. The Feynman diagrams of $\gamma\gamma 
\rightarrow \tilde\nu + l^{-} l^{'+}$ are presented in Fig.1,
and the Feynman diagrams of $\gamma\gamma
\rightarrow \tilde\nu + q \bar{q}^{'}$ will be similar.
We can calculate the cross sections of the processes from these diagrams 
and then fold the cross sections with photon luminosity to get observable 
results in $e^{+}e^{-}$ collider.

We also need to consider the decay channels of sneutrinos
in order to discuss the experimental detection possibilities.
There are two essentially different modes in our case for
sneutrino decay: it may be the LSP, in which case the $\rlap /\! R_{p}$
decays are unique, or it may decay to some of the neutralinos or 
charginos if they are lighter than the sneutrino.

If sneutrino is the  LSP, it will decay through $R_p$-violating
terms.
With nonzero
$\lambda$ coupling, sneutrino will decay to two leptons 
and with
nonvanishing $\lambda^{'}$ coupling sneutrino will
decay to a quark pair. 
If we assume that $\lambda$ or  $\lambda^{'}$
including different flavours is nonzero, sneutrinos may even
decay to different flavours of a lepton pair or a quark pair.
Even in the flavour conserving case the background from Higgs is 
negligible.
In order to distinguish the signal from a Z boson decaying
to two leptons, we have to consider the different spins of
sneutrinos and Z boson and the subsequent angular distributions.
When $m_{\tilde{v}}$ is not close to $m_z$, invariant mass cut
may also be useful.

If the lighter neutralinos $\tilde{\chi}^{0}_{1,2}$ and the lighter
chargino $\tilde{\chi}^{\pm}_{1}$ are lighter than the  
sneutrinos,
then the $R_p$-conserving decay is possible.
The possible decay channels are as follows:
$$
\begin{array}{lll}
\tilde{\nu}_i \rightarrow \tilde{\chi}_1^{\pm} l^{\mp}_i,\;
\tilde{\nu}_i \rightarrow \tilde{\chi}_{1,2}^{0} \nu _i.
\end{array}
\eqno(2.2)
$$
If it is kinematically allowed, sneutrino can also decay as follows
$$
\begin{array}{lll}
\tilde{\nu}_i \rightarrow \tilde{l}^{\pm}_{iL} W^{\mp}.
\end{array}
\eqno(2.3)
$$
In the rest of this section, we will 
assume the GUT relations between the $SU(2)$ and the $U(1)$ gaugino mass
parameters, namely
$$
\begin{array}{l}   
M_2 = {\alpha_2 \over \alpha_3} {m_{\tilde g}} \\
\\
M_1 = {5 \over 3} {{\tan}^2 {\theta_W}} {M_2},
\end{array}   
\eqno(2.4)
$$
where $m_{\tilde g}$ is the mass of the gluino.

We take as a representative point in the MSSM parameter space 
the following:
$\mu = 500$ GeV, $\tan\beta = 10$, $m_{\tilde g} = 300$ GeV.
The corresponding chargino and light neutralino  masses used in our
numerical calculations are
as follows: $m_{\chi_1^0}\sim 42$ GeV, $m_{\chi_2^0}\sim 82$ GeV,
$m_{\chi_1^\pm}\sim 81$ GeV, and $m_{\chi_2^\pm}\sim 513$ GeV.
In our case 
only the decays in (2.2) are allowed.
In Fig.2 we show the branching ratios of the decays of
sneutrinos with parameter $\lambda$ or
$\lambda^{'}$ dominating.
 
We can see from the Fig. 2 that the $R_p$-violating decay
of sneutrinos will be important if we take $\lambda_{131}=0.062$,
and even dominate when $\lambda_{322}^{'} =0.52$.
In the case of flavour-changing coupling, the signal events 
would be easier to detect. 
In the following we use an $R_p$ violating decay of 
$\tilde{\nu}$ as a main way  to detect the process (see Fig. 3 and Fig. 4), 
which means that the signal events have four leptons or four jets
with invariant-mass equal to the C.M. energy of the $\gamma \gamma$ 
collision.

However, we also need consider other decay modes if the $R_p$-violating
parameters are small. 
In Fig.5 we have plotted the production cross section of
sneutrinos with sneutrino decaying to $\chi^{+}+l^{-}$  
with $\lambda \sim 0.01$. In that 
case the final states include mainly four leptons and missing energy.

\begin{flushleft} {\bf 3. Numerical results} \end{flushleft}

\noindent
In our numerical calculations, we take the single-coupling
assumption: only one $\lambda$ or $\lambda^{'}$ coupling
dominates at a time.  

 In Fig.3(a), we show the cross section of $
\gamma\gamma (\rightarrow \tilde{\nu}_{\tau}+ e^{+} e^{-} )
\rightarrow  e^{+} e^{-} e^{+} e^{-}$ as a function of 
mass of sneutrino $\tilde{\nu}_{\tau}$. 
We plot the Figure with $\sqrt{s_{ee}}=500$ GeV (dotted line) and
with $\sqrt{s_{ee}}=1$ TeV (solid line).
The values of the couplings used, $\lambda_{131}=0.062$ and 
$\lambda_{131}=0.03$ are denoted in the Figure. 
The cross section may be of the order of 0.1 fb for light $m_{\tilde{\nu}}$ 
if we take $\lambda_{131}=0.062$ (present upper limit for 
$m_{\tilde e_R}=100$ GeV). Even 
for an $R_p$-violating coupling $\lambda_{131}=0.03$, 
one signal event with four leptons may be produced with  
$m_{\tilde{\nu}}=110$ 
GeV\footnote{The present lower limits for sneutrino
masses are $m_{\tilde\nu_2}>84$ GeV and $m_{\tilde\nu_3}>86$ GeV
\cite{lepc}.} at LC with an integrated luminosity 500 $fb^{-1}$.
Similarly, in Fig.3(b), we plot the
cross section of $\gamma\gamma 
(\rightarrow \tilde{\nu}_{\tau} + e^{+} \mu^{-})
\rightarrow e^{+} e^{+} \mu^{-}\mu^{-}$ as a function of masses of 
sneutrino $\tilde{\nu}_{\tau}$ with
the coupling of $\lambda$, where $\lambda_{231}=0.07$ (present
limit for $m_{\tilde e_R}=100$ GeV). 
For light $\tilde{\nu}$, about 20 flavour-changing
events may be produced with 500 fb$^{-1}$ integrated luminosity.

In Fig.4 (a), we plot the cross section of $\gamma\gamma (\rightarrow
\tilde{\nu}_{\tau} + s \bar{s}) \rightarrow s s \bar{s} \bar{s}$ as a 
function of sneutrino mass $m_{\tilde{\nu}_{\tau}}$ with 
$\lambda^{'}_{322}=0.52$, and similarly $ \gamma\gamma 
(\rightarrow \tilde{\nu}_{\tau} b \bar{s}) \rightarrow b b 
\bar{s}\bar{s}$
with $\lambda^{'}_{323}=0.52$ is plotted in Fig.4 (b). It
is seen that with an integrated luminosity 500 $fb^{-1}$  hundreds of
signal events are produced if $m_{\tilde{\nu}_{\tau}} \lsim 200$ GeV
for $ss\bar{s}\bar{s}$. 
Also hundreds
of flavour-changing signals ($bb\bar{s}\bar{s}$) for light
$\tilde{\nu}_{\tau}$ are produced. 

In our parton level Monte Carlo analysis we have used an angular cut of 
$5^0 < \theta < 175^0$ for the leptons/jets in the final states. A minimum
energy cut of 1.3 GeV has  been applied to the leptons/jets in the 
final states.

In Fig.5(a) and (b) the cross sections of $\gamma\gamma (\rightarrow
\tilde{\nu}_{\tau} + e^{+}e^{-})
\rightarrow \tilde{\chi}_{1}^{+} \tau^{-} e^{+}e^{-}$ and
$\gamma\gamma (\rightarrow
\tilde{\nu}_{\tau} + e^{+}\mu^{-})
\rightarrow \tilde{\chi}_{1}^{+} \tau^{-} e^{+}\mu^{-}$, 
with $\lambda_{131}=0.01$ and $\lambda_{231}=0.01$ respectively,
are presented. We find that it may be possible to detect single
sneutrino events even for smaller $R_p$ violating
coupling, where signals come mainly from $R_p$ conserving
decay. 

Comparing with the results of single sneutrino resonant production
\cite{RP_spp2}, we find that the cross sections of sneutrino produced in 
association with two leptons or two jets are of similar magnitude and 
can even be larger. 
Especially for $R_p$ violating
coupling involving light flavours, this kind of processes involving a 
single 
sneutrino with two leptons or jets can be easily detected. 

\begin{flushleft} {\bf 4. Conclusion} \end{flushleft}

\noindent
We have studied the single sneutrino production (accompanied 
by two leptons or two 
jets) and decay of sneutrino in $\gamma\gamma$ collisions, in the 
context of R-parity violating supersymmetry.
The cross section for the processes, in the future LC experiments with 
$e^+e^-$ C.M. energy 1 TeV,  with sneutrino mass below 175 GeV is above 
0.01 fb with $\lambda=0.062$, 
and with sneutrino mass below 200 GeV is above
1 fb with $\lambda^{'}=0.52$, allowed by 
experimental limits. Even for much smaller $R_p$ violating couplings, 
sneutrinos may be detected with their $R_p$ conserving decay mode. 
We can also detect the flavour-changing final states if relevant $R_p$ 
violating couplings are close to the present experimental 
limits.  If we cannot find any such signals from the experiments, we could 
improve upon the present upper bounds on $\lambda$ and $\lambda^{'}$ or 
increase the lower limit on sneutrino mass.

\begin{flushleft} {\bf Acknowledgements} \end{flushleft}

\noindent
The authors thank the Academy of Finland
(project numbers 163394 and 48787) for financial support.
Z.-H. Yu thanks the World Laboratory, Lausanne, for the scholarship. 
S.R. wishes to acknowledge the hospitality provided by the Helsinki 
Institute of Physics where this work was done. 
The authors thank Probir Roy for pointing out a missing graph in the
previous version.

\newpage
\begin{center} Appendix\end{center}
$\gamma\gamma$ collision
\par
  In order to get the observable results in the measurements of
snutrino production via $\gamma \gamma$ fusion
in $e^{+}e^{-}$ collider, we need to fold the cross section of
$\gamma\gamma
\rightarrow \tilde{\nu} + l \bar{l}^{'} (q \bar{q}^{'})$ with the photon 
luminosity,
$$
\sigma(s) = \int_{m_{\phi^{'}}/\sqrt{s}}^{x_{max}} dz
\frac{dL_{\gamma\gamma}}{dz}
              \hat{\sigma}(\hat{s}),
\eqno {(A.1)}
$$
where $\hat{s}=z^2 s$, $\sqrt{s}$ and $\sqrt{\hat{s}}$ are the
$e^{+}e^{-}$
and $\gamma\gamma$ c.m. energies respectively, and
$\frac{dL_{\gamma\gamma}}{dz}$
is the photon luminosity, which is defined as \cite{pho_tech}
$$
\frac{dL_{\gamma\gamma}}{dz} = 2z \int_{z^{2}/x_{max}}^{x_{max}}
\frac{dx}{x}
                                F_{\gamma /e}(x)F_{\gamma /e}(z^{2}/x).
\eqno {(A.2)}
$$
The energy spectrum of the back-scattered photon is given by 
\cite{pho_tech}.
$$
F_{\gamma /e}(x) = \frac{1}{D(\xi)}[1-x+\frac{1}{1-x}-\frac{4x}{\xi
(1-x)}+
                   \frac{4x^{2}}{\xi^{2} (1-x)^{2}}].
\eqno {(A.3)}
$$
taking the parameters of Ref. \cite{PP_Lum}, we have
$\xi=4.8$, $x_{max}=0.83$ and $D(\xi)=1.8$.
  
\begin{figure}[b]
\leavevmode
\begin{center}
\mbox{\epsfxsize=8.truecm\epsfysize=5.truecm\epsffile{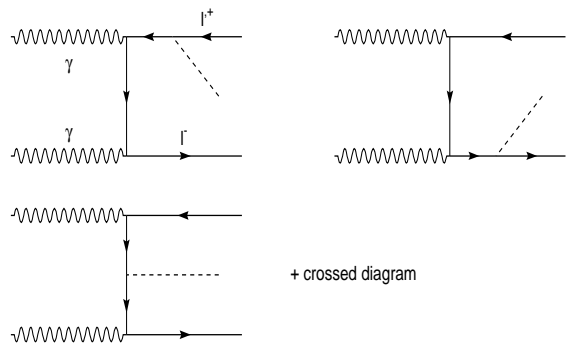}}
\end{center}
\caption{\label{fig1}
         Feynman diagrams of $\gamma\gamma\rightarrow \tilde{\nu}_{\mu}
         l^{-} l^{'+}$. Dashed line represents sneutrino.}
\end{figure}

\begin{figure}[t]
\leavevmode
\begin{center} 
\mbox{\epsfxsize=6.truecm\epsfysize=6.truecm\epsffile{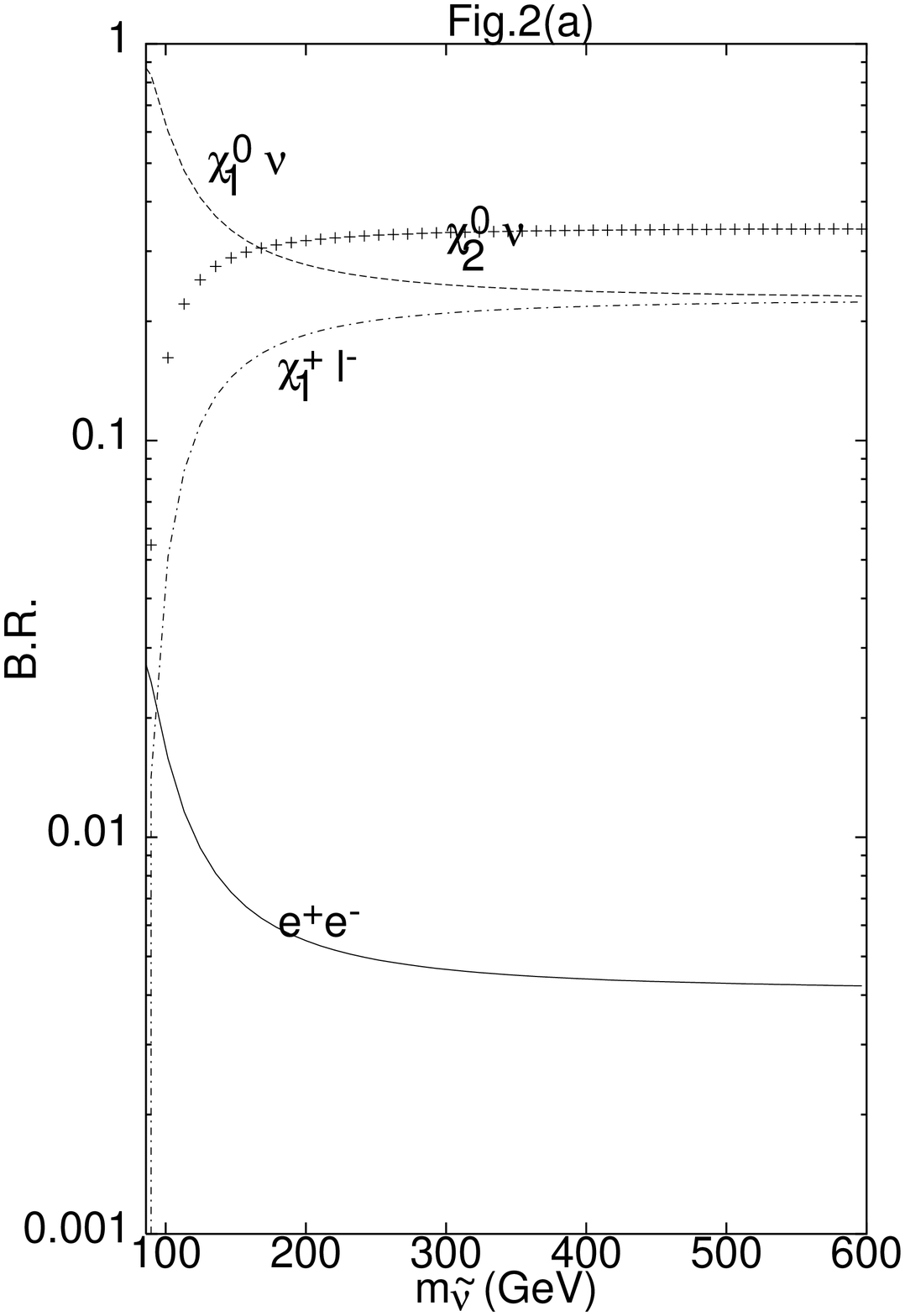}}
\mbox{\epsfxsize=6.truecm\epsfysize=6.truecm\epsffile{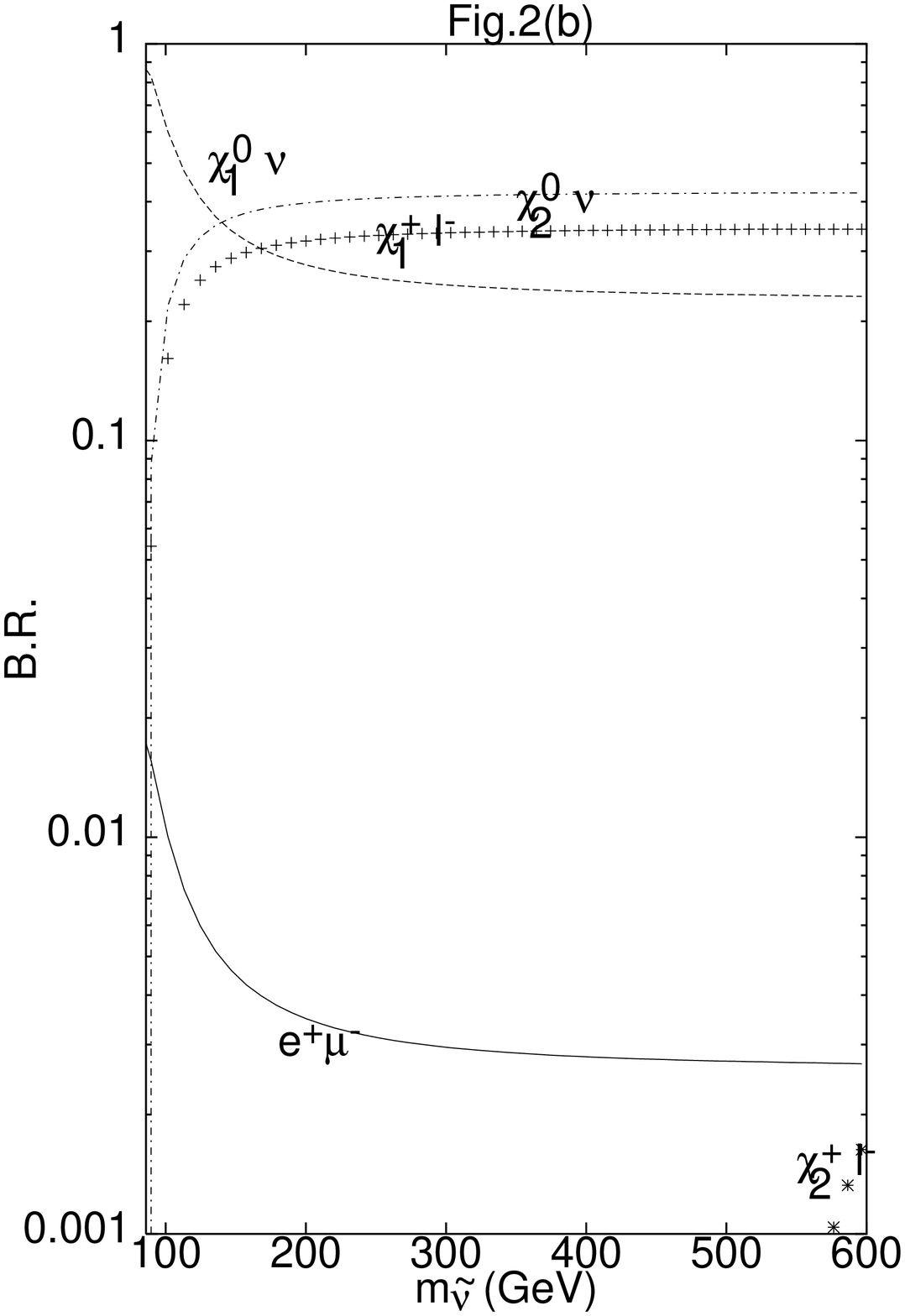}}
\vspace*{1.truecm}
\\
\mbox{\epsfxsize=6.truecm\epsfysize=6.truecm\epsffile{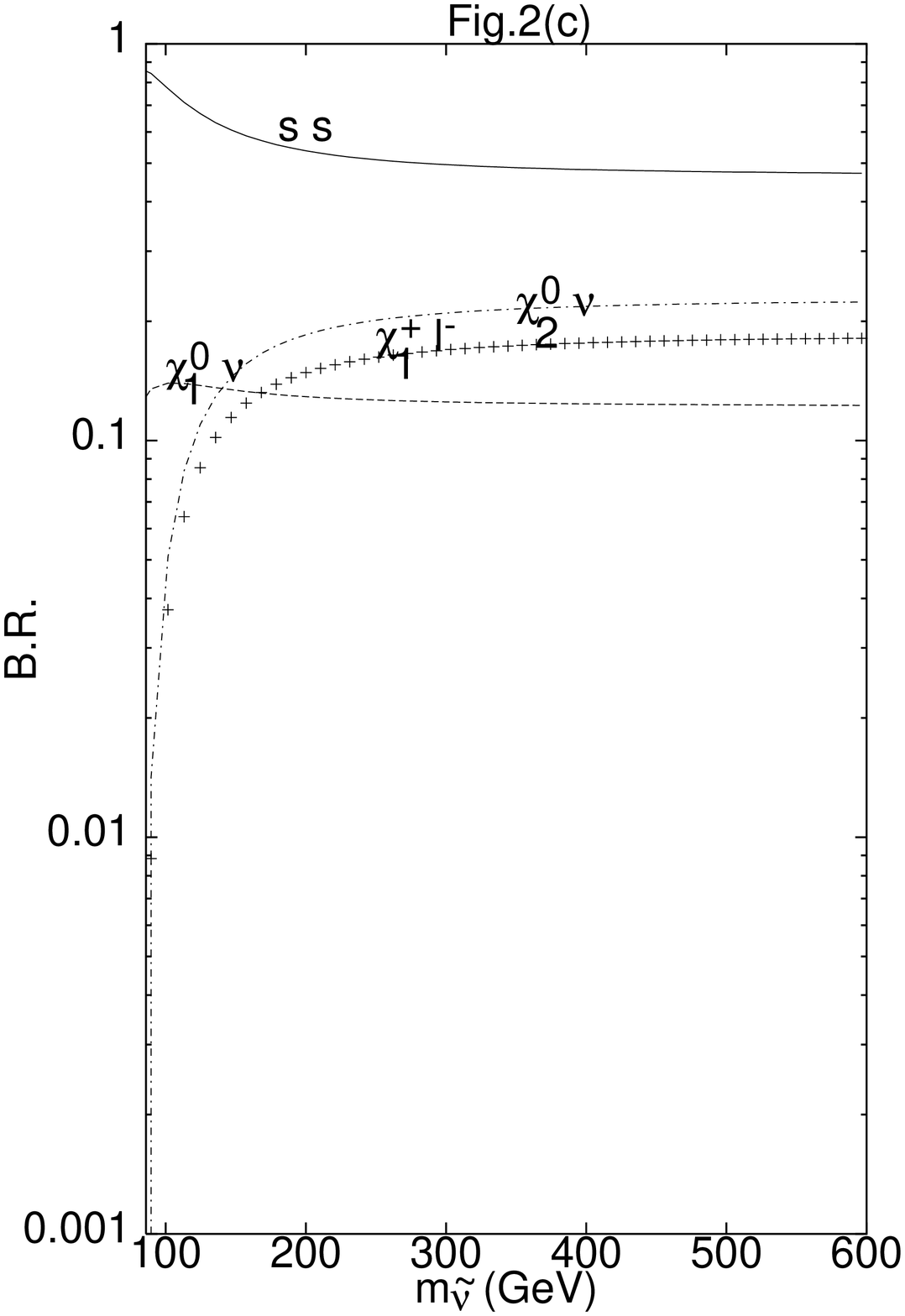}}
\mbox{\epsfxsize=6.truecm\epsfysize=6.truecm\epsffile{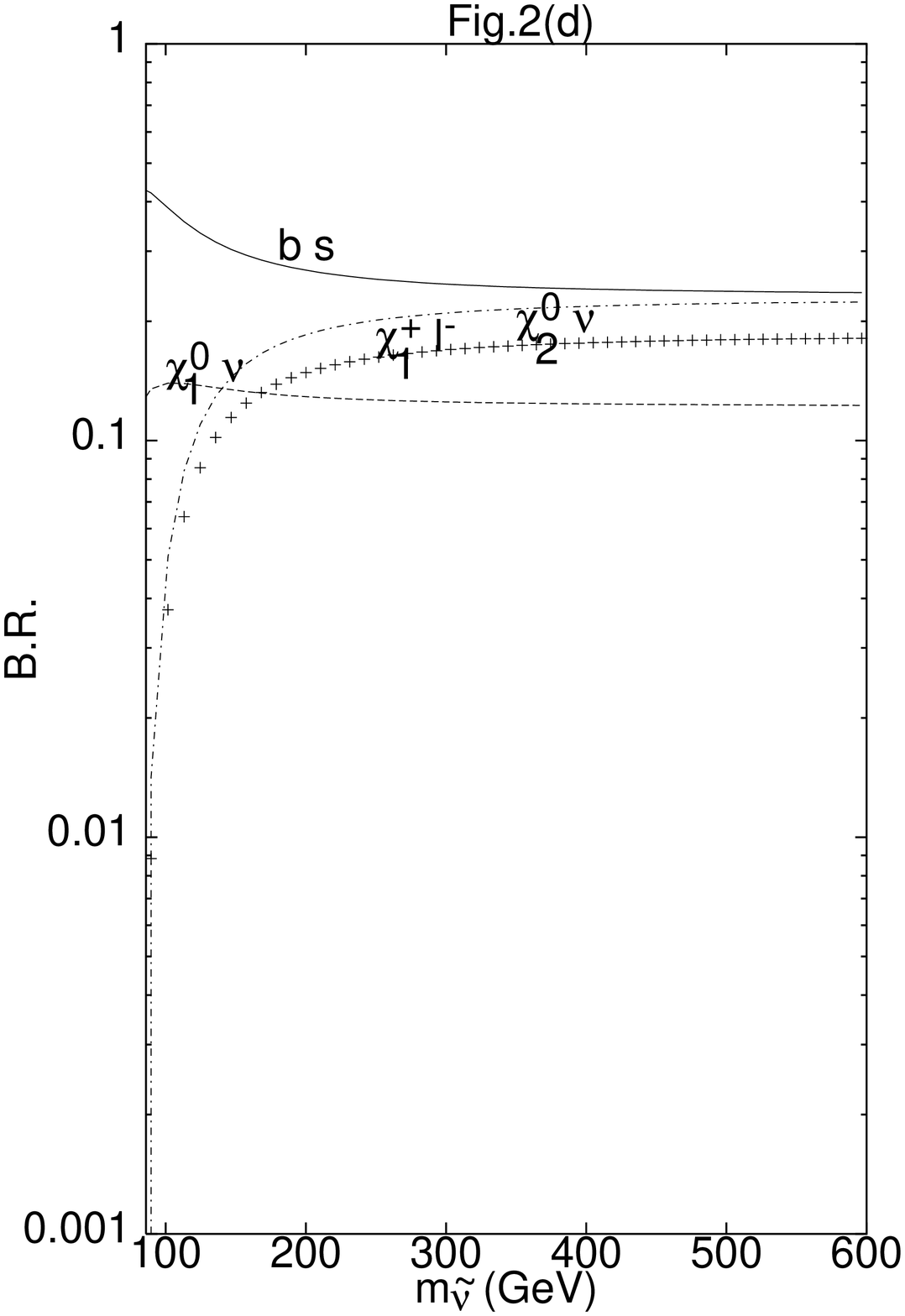}}
\end{center}
\caption{\label{fig2}
Branching ratios for sneutrino $\tilde{\nu}_{\tau}$ decays
        as a function of mass of sneutrino $\tilde{\nu}_{\tau}$,
(a)     with $\lambda_{131}=0.062$,
(b)     with $\lambda_{231}=0.07$,
(c)     with $\lambda^{'}_{322}=0.52$, and
(d)     with $\lambda^{'}_{323}=0.52$.}
\end{figure}

\begin{figure}[t]
\leavevmode
\begin{center}
\mbox{\epsfxsize=6.truecm\epsfysize=6.truecm\epsffile{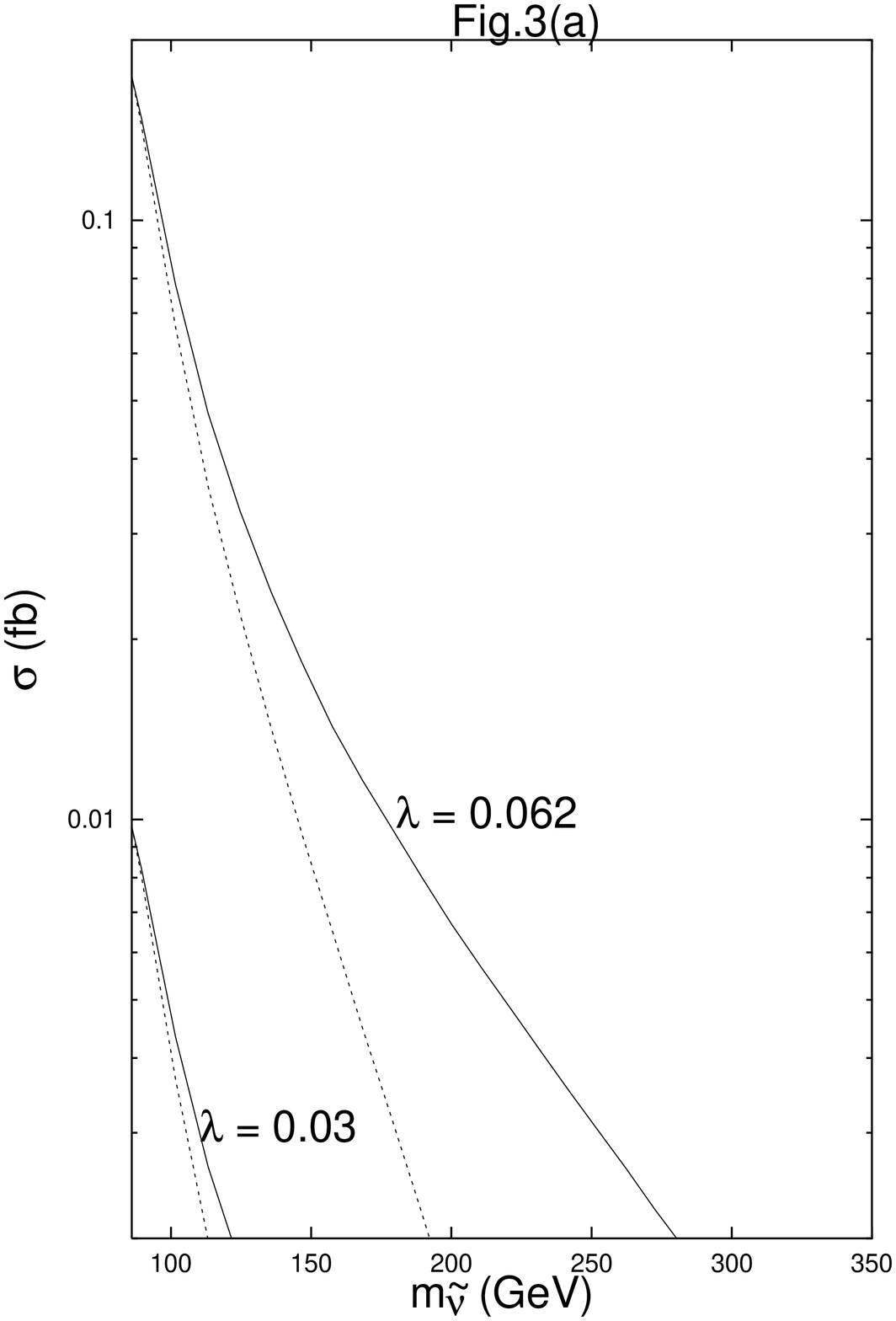}}
\mbox{\epsfxsize=6.truecm\epsfysize=6.truecm\epsffile{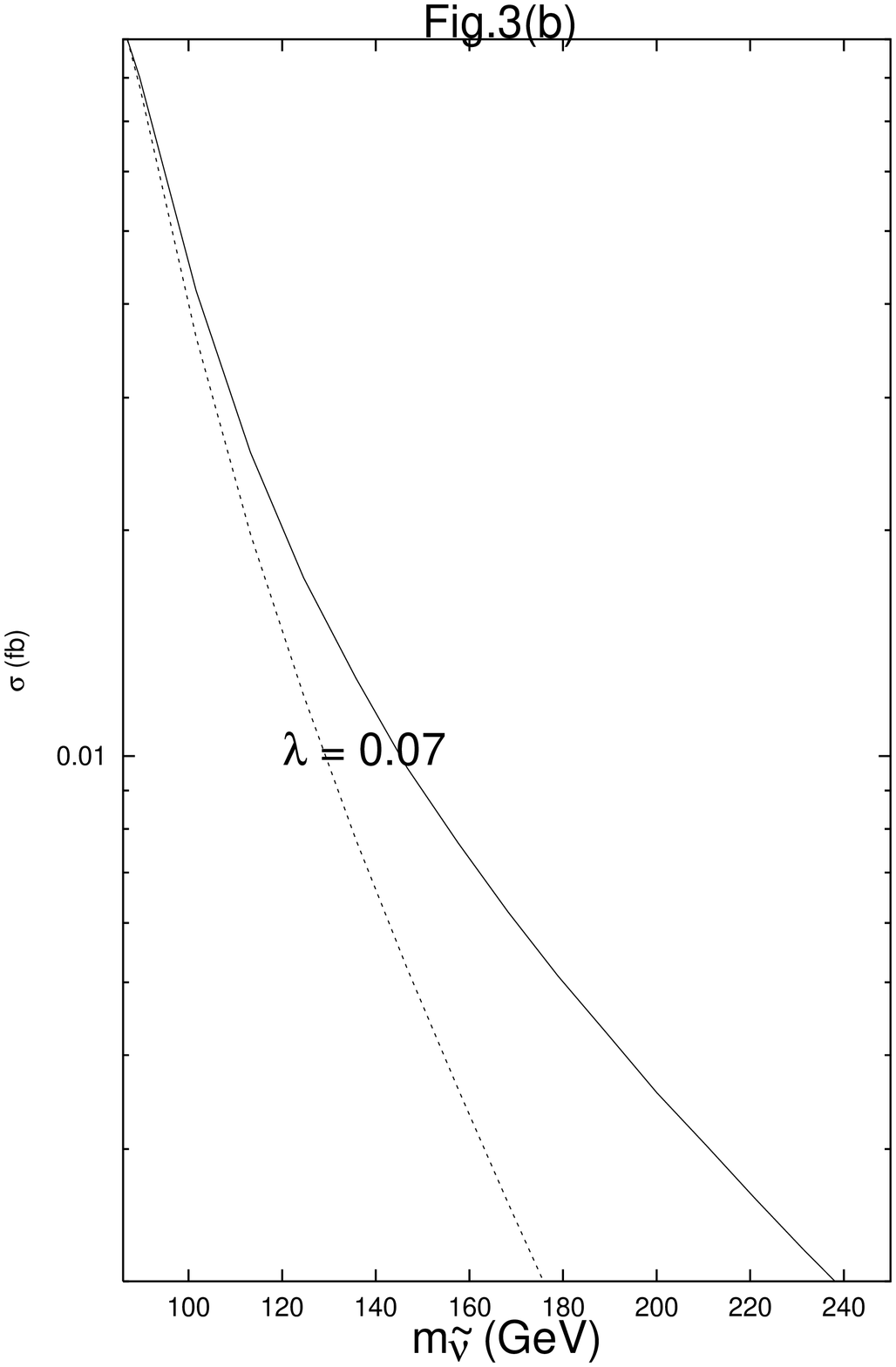}}
\end{center}
\caption{\label{fig3}
Cross section (a) of $\gamma\gamma
        (\rightarrow \tilde{\nu}_{\tau}+ e^{+}e^{-})
        \rightarrow e^{+}e^{+}e^{-}e^{-} $ with $\lambda_{131}=0.062$
and  $\lambda_{131}=0.03$,
and (b) of $\gamma\gamma
        (\rightarrow \tilde{\nu}_{\tau}+ e^{+}\mu^{-})
        \rightarrow e^{+}e^{+}\mu^{-}\mu^{-} $ with
$\lambda_{231}=0.07$.
Both are as a function of mass of 
        sneutrino $\tilde{\nu}_{\tau}$.   
Solid lines and the dotted lines correspond to the $e^+e^-$ C.M. energy 
1 TeV and 500 GeV respectively.}
\end{figure}
\begin{figure}[t]
\leavevmode
\begin{center}
\mbox{\epsfxsize=6.truecm\epsfysize=6.truecm\epsffile{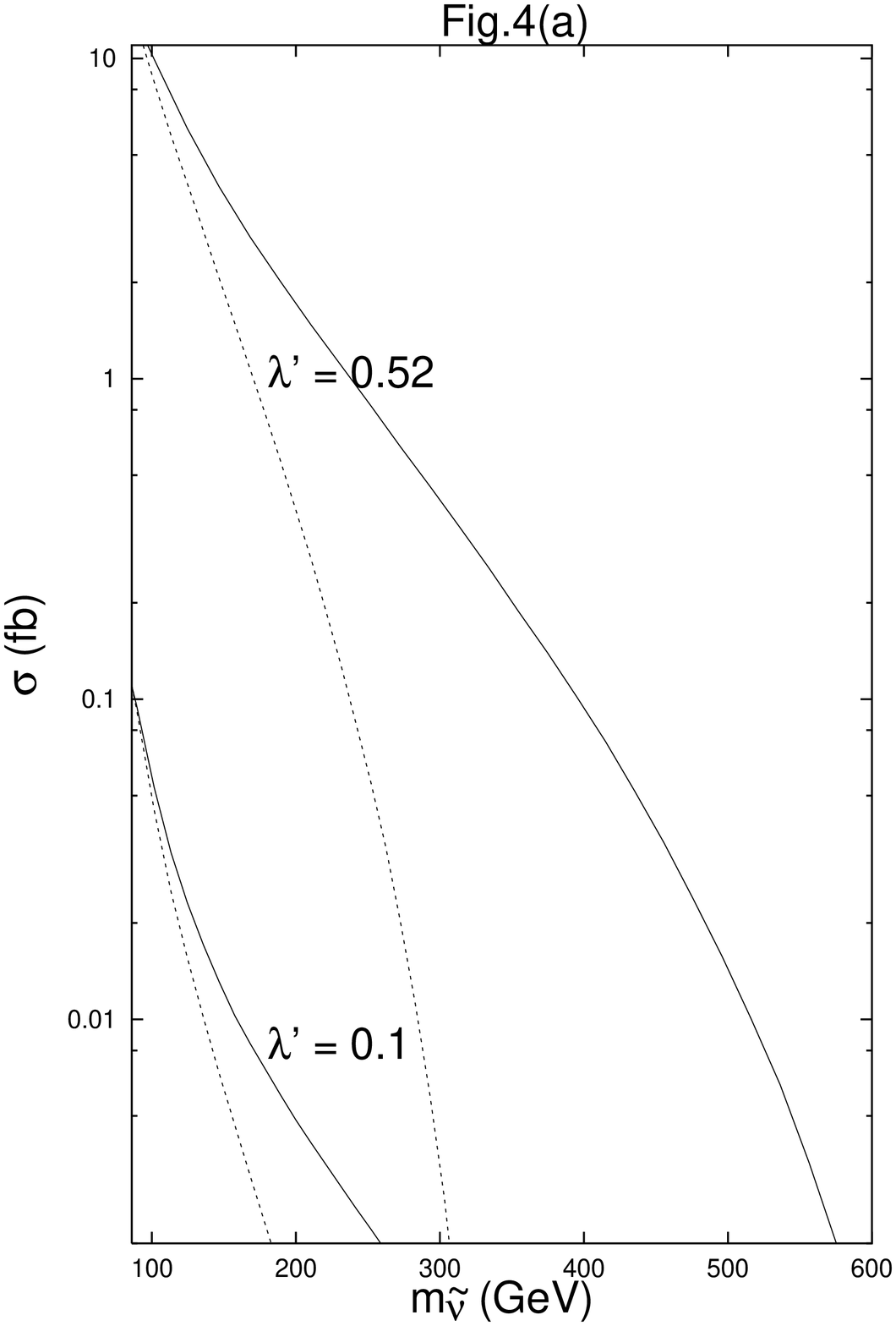}}
\mbox{\epsfxsize=6.truecm\epsfysize=6.truecm\epsffile{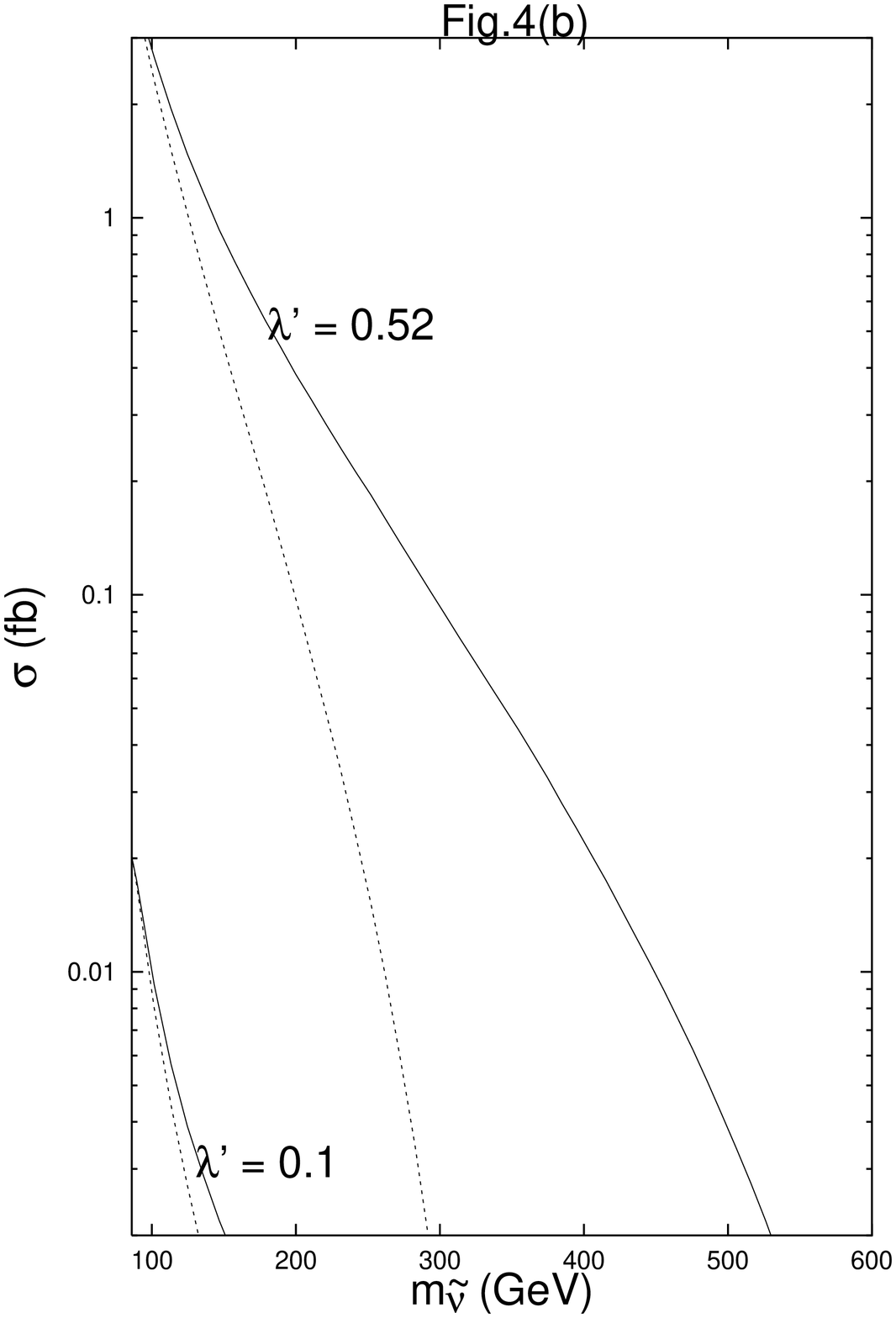}}
\end{center}
\caption{\label{fig4}     
Cross section (a) of $\gamma\gamma
        (\rightarrow \tilde{\nu}_{\tau} + s \bar{s})  
        \rightarrow s s \bar{s}\bar{s} $ and (b)
 of $\gamma\gamma
        (\rightarrow \tilde{\nu}_{\tau}+ b\bar{s})
        \rightarrow bb\bar{s}\bar{s} $, as a function of mass of
        sneutrino $\tilde{\nu}_{\tau}$, and with
        $\lambda^{'}_{323}=0.52$ and
        $\lambda^{'}_{323}=0.1$.
Solid lines and the dotted lines correspond to the $e^+e^-$ C.M. energy
1 TeV and 500 GeV respectively.}
\end{figure}

\begin{figure}[t]
\leavevmode   
\begin{center}
\mbox{\epsfxsize=6.truecm\epsfysize=6.truecm\epsffile{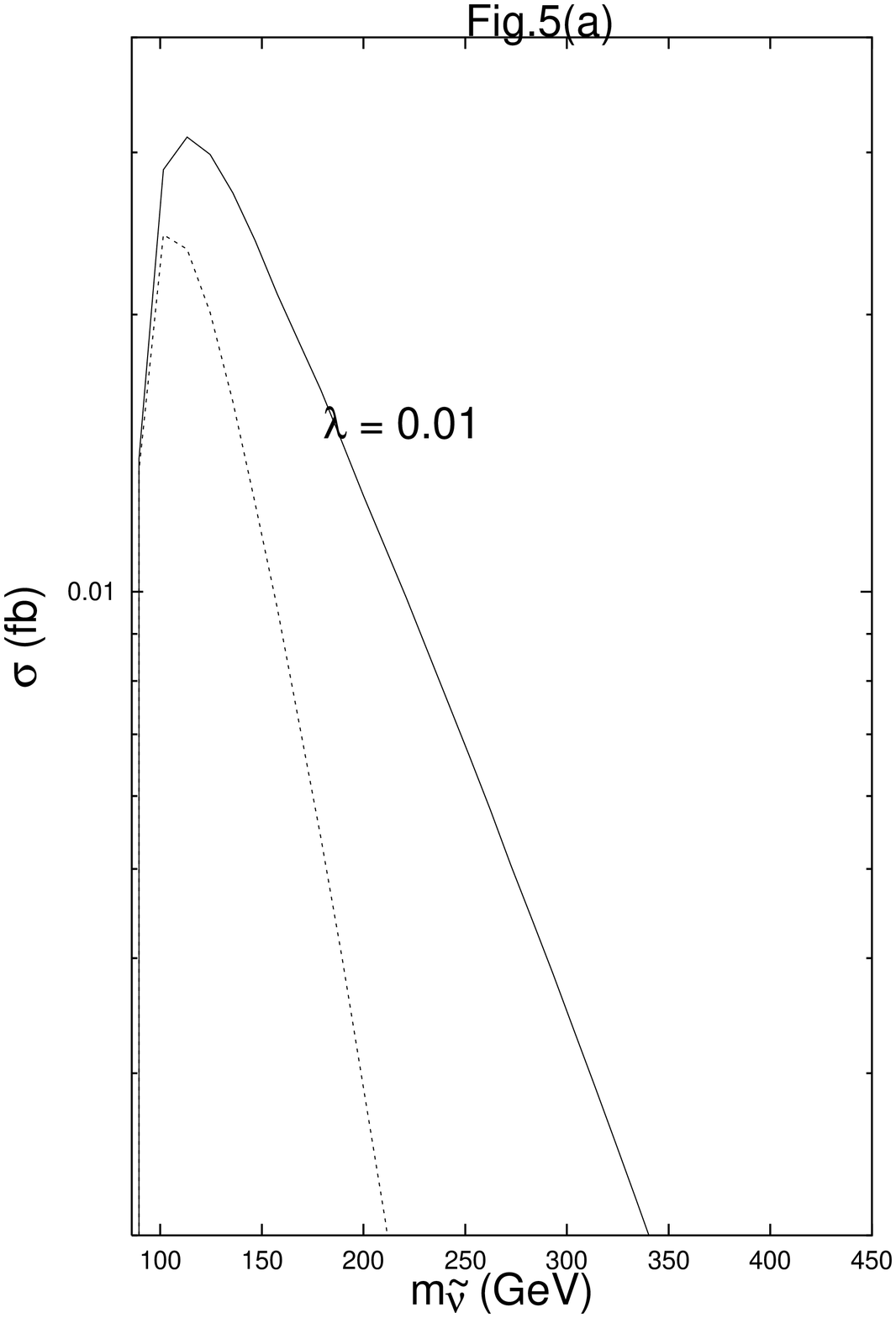}}
\mbox{\epsfxsize=6.truecm\epsfysize=6.truecm\epsffile{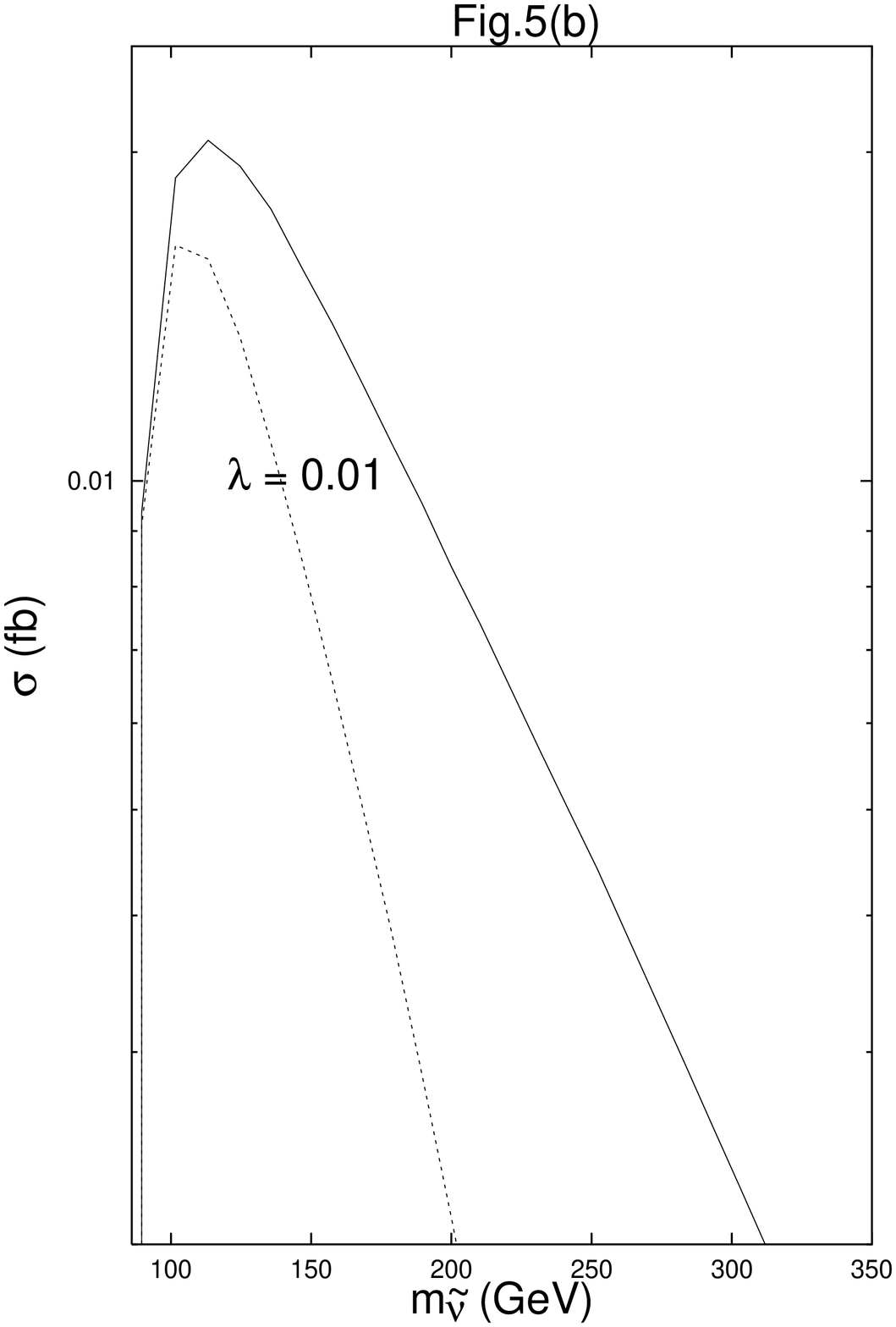}}
\end{center}
\caption{\label{fig5}
Cross section (a) of $\gamma\gamma
        \rightarrow \tilde{\chi}_{1}^{+} \tau^{-} e^{+}e^{-}$
with $\lambda_{131}=0.01$, and (b) of $\gamma\gamma
        \rightarrow \tilde{\chi}_{1}^{+} \tau^{-} e^{+}\mu^{-}$
with $\lambda_{231}=0.01$.  Both are
         as a function of mass of sneutrino $\tilde{\nu}_{\tau}$.
Solid lines and the dotted lines correspond to the $e^+e^-$ C.M. energy
1 TeV and 500 GeV respectively.}
\end{figure}


\newpage
\begin{thebibliography}{s20}

\bibitem{RP_Farrar} G.Farrar and P.Fayet, Phys. Lett. {\bf B76} (1978) 
             575.

\bibitem{RP_Hall} L.J. Hall and M. Suzuki, Nucl. Phys. {\bf B231} (1984) 
419.

\bibitem{RP_Aulakh} C.S. Aulakh, R.N. Mohapatra, Phys. Lett. {\bf 119B}
            (1982) 136;
             J. Ellis, G. Gelmini, C. Jarlskog, G.G. Ross and
             J.W.F. Valle, Phys. Lett. {\bf 150B }(1985) 142;
             G.G Ross and J.W.F. Valle, Phys. Lett. {\bf 151B }(1985) 375;
             A. Santamaria and J.W.F. Valle, Phys. Lett. {\bf195B }(1987) 423.

\bibitem{neutrino} Y. Fukuda et al, Phys. Rev. Lett., {\bf 81 }(1998) 
1562.

\bibitem{SUSY} H.E. Haber and G.L. Kane, Phys. Rep. {\bf 117} (1985) 75;
             J.F. Gunion and H.E. Haber, Nucl. Phys. {\bf B272} (1986) 1.

\bibitem{pho_tech} V.Telnov, Nucl. Instrum. Meth. {\bf A294} (1990) 72; L. 
Ginzburg,  
             G. Kotkin and H. Spiesberger, Fortschr. Phys. {\bf 34}
             (1986) 687.

\bibitem{RP_Phe} S.Weinberg, Phys. Rev. {\bf D26 }(1982) 287;
              P. Roy, TIFR/TH/97-60;
              K. Agashe and M. Graesser, Phys. Rev. {\bf D54 }(1996)
             4445;
             M. Chaichian, K. Huitu, Phys.Lett. {\bf B384 }(1996) 157;
             K. Huitu, J. Maalampi, M.Raidal and A.Santamaria,
             Phys.Lett. {\bf B430 }(1998) 355;
             J-H. Jiang, J.G. Kim and J.S. Lee, Phys. Rev.
             {\bf D55 }(1997) 7296;
             Phys. Lett. {\bf B408} (1997) 367; Phys. Rev.
             {\bf D58 }(1998) 035006;
             G. Bhattacharyya, D. Choudhury and K. Sridhar,
             Phys. Lett. {\bf B355 }(1995) 193;
             J. Ferrandis, Phys. Rev. {\bf D60 }(1999) 095012;
             K. Huitu, K. Puolam\"aki, D.-X. Zhang, Phys.Lett. 
             {\bf B446 }(1999) 285;
             M. A. Diaz, J. Ferrandis, J. C. Romao and J. W. F. Valle,
             Nucl. Phys. {\bf B590} (2000) 3;
             M. A. Diaz, J. Ferrandis and J. W. F. Valle, 
             Nucl. Phys. {\bf B573} (2000) 75;
             B. C. Allanach, A. Dedes and H. K. Dreiner, Phys. Rev. {\bf
             D60 }(1999) 075014;
             S. Bar-Shalom, G. Eilam and J. Wudka,
             Phys.Rev. {\bf D59 }(1999) 035010;
             S. Bar-Shalom, G. Eilam and A. Soni,
             Phys.Rev. {\bf D60 }(1999) 035007.

\bibitem{RP_spp}S.Dimopoulos and L.J. Hall, Phys. Lett. {\bf B207} (1987) 
210;
             V.Barger, G.F. Giudice and T. Han, Phys. Rev. {\bf D40}
            (1989) 2987; M. Czakon and J. Gluza, hep-ph/0003228.

\bibitem{RP_decay}J. Erler, J.L. Feng and N. Polonsky, Phys. Rev. Lett.
            {\bf 78} (1997) 3063; D.K. Ghosh and S. Roy,  Phys. Rev. {\bf D63} 
            (2001) 055005; D. Choudhury and A. Datta, Nucl. Phys. {\bf B592} 
            (2001) 35.

\bibitem{RP_indirect}D. Choudhury, Phys. Lett. {\bf B376} (1996) 201;
            D.K. Ghosh, S. Raychaudhuri and K. Sridhar,
            Phys. Lett. {\bf B396} (1997) 177; M. Chemtob and
            G. Moreau, Phys. Rev. {\bf D59} (1999) 116012;
            Z-H. Yu, H. Pietschmann, W-G. Ma, L. Han and Y. Jiang,
            Eur. Phys. J. {\bf C16 }(2000) 541;
            Eur. Phys. J. {\bf C16 }(2000) 695.
             
\bibitem{RP_spp2}M. Chaichian, K. Huitu and Z.-H. Yu, hep-ph/0101272;
		 X. Yin, W.-G. Ma, L.-H. Wan, Y. Jiang and L. Han, 
                 hep-ph/0106183.

\bibitem{RP_BV}G.F. Deliot, C. Royon, E. Perez, G. Moreau and M. Chemtob,
             Phys. Lett. {\bf B475} (2000) 184;
             G.Moreau, E.Perez and G.Polesello, hep-ph/0003012.

\bibitem{bgh}V. Barger, G.F. Giudice, T. Han, Phys. Rev. {\bf D40}
             (1989) 2987.

\bibitem{ls} F. Ledroit, G. Sajot, GDR-S-008 (ISN, Grenoble, 1998),
   http://qcd.th.u-psud.fr/GDR\_SUSY/GDR\_SUSY\_PUBLIC/entete\_note\_publique.

\bibitem{RP_V_form}R. Barbier et.al, hep-ph/9810232; B. Allanach et. al,
            hep-ph/9906224.

\bibitem{bilinear} L.~J.~Hall and M.~ Suzuki, Nucl. Phys. {\bf B231}
           (1984) 419 ; I-H.~Lee, Phys. Lett. {\bf B138} (1984) 121; 
           Nucl. Phys. {\bf B246} (1984) 120;
           S.~Dawson, Nucl. Phys. {\bf B261} (1985) 297;
           F.~de Campos {\it et al.}, Nucl. Phys. {\bf B451} (1995) 3;
           M.~Nowakowski and A.~Pilaftsis, Nucl. Phys. {\bf B461} (1996) 
           19; J.~W.~F.~Valle, hep-ph/9808292; 
           R.~Hempfling, Nucl. Phys. {\bf B478} (1996) 3;
           H.~P.~Nilles and N.~Polonsky, Nucl. Phys. {\bf B484} (1997) 
           33; B.~de Carlos and P.~L.~White, Phys. Rev. {\bf D55} (1997) 
           4222; E.~Nardi, Phys. Rev. {\bf D55} (1997) 5772;
           S.~Roy and B.~Mukhopadhyaya, Phys. Rev. {\bf D55} (1997) 7020;
           M.~A.~D\'\i az {\it et al}, Nucl. Phys. {\bf B524} (1998) 23;
           E.~J.~Chun {\it et al.}, Nucl. Phys. {\bf B544} (1999) 89;
           M.~Hirsch {\it et al.}, hep-ph/0004115; V.~Bednyakov, 
           A. Faessler and S. Kovalenko, Phys. Lett. {\bf B442} (1998) 
           203; B.~Mukhopadhyaya, S. Roy and F. Vissani,  Phys. Lett. 
           {\bf B443} (1998) 191; A. Datta, B. Mukhopadhyaya and 
           S. Roy, Phys. Rev. {\bf D61} (2000) 055006.


\bibitem{PP_Lum}K.Cheung, Phys. Rev. {\bf D47} (1993) 3750; ibid. 50 
(1994) 1173.

\bibitem{lepc} C. Mariotti (DELPHI Collaboration) reports to the Open
Session of the LEP Experiments Committee on 7 March, 2000, available
from http://www.cern.ch/~offline/physics\_links/lepc.html.
\end{thebibliography}
\end{document}